\documentclass[ALICE,manyauthors]{cernphprep}
\usepackage[comma,square,numbers,sort&compress]{natbib}
\usepackage{hyperref} 
\usepackage{lineno}
\usepackage{color}
\usepackage{multirow}

\usepackage{graphicx}
\usepackage{epstopdf}
\usepackage{amssymb}
\usepackage{xspace}
\usepackage{amsmath}
\usepackage{upgreek}
\usepackage{hyperref}
\usepackage{doi}
\usepackage{etoolbox}
\usepackage[normalem]{ulem}  
\usepackage[T1]{fontenc}

\patchcmd{\thebibliography}{\section*{\refname}}{}{}{}

\newcommand{\snn}{$\sqrt{s_{\mathrm{NN}}}$}
\newcommand{\s}{$\sqrt{s}$}

\newcommand{\kst}{$K^{*0}$}

\newcommand{\kstba}{$\overline{K}^{*0}$}

\newcommand{\ph} {$\phi$}

\newcommand{\pt}{$p_{\mathrm{T}}$}

\newcommand{\Kzs}{\mbox{$K^0_S$}}
\newcommand{\pb}{\mbox{Pb--Pb}}

\newcommand{\rh}{\ensuremath{\mathrm{\rho_{00}}}}

\usepackage{float} 

\begin{document}
\begin{titlepage}

\PHyear{2019}
\PHnumber{251}      
\PHdate{30 October}  
  

\title{Evidence of spin-orbital angular momentum interactions in relativistic heavy-ion collisions}   
\ShortTitle{Spin alignment of vector mesons}
\Collaboration{ALICE Collaboration\thanks{See Appendix~\ref{app:collab} for the list of collaboration members}}
\ShortAuthor{ALICE Collaboration} 

\begin{abstract}
 The first evidence of spin alignment of vector mesons (\kst~and \ph) in heavy-ion collisions at the Large Hadron Collider (LHC) is reported. The spin density matrix element \rh~is measured at midrapidity ($|y|~<$ 0.5) in Pb--Pb collisions at a center-of-mass energy (\snn) of 2.76 TeV with the ALICE detector. \rh~values are found to be less than 1/3 (1/3 implies no spin alignment) at low transverse momentum (\pt~$<$ 2 GeV/$c$) for \kst~and \ph~at a level of 3$\sigma$ and 2$\sigma$, respectively. No significant spin alignment is observed for the \Kzs~meson (spin = 0) in Pb--Pb collisions and for the  vector mesons in $pp$ collisions. The measured spin alignment is unexpectedly large but qualitatively consistent with the expectation from models which attribute it to a polarization of quarks in the presence of angular momentum in heavy-ion collisions and a subsequent hadronization by the process of recombination.
\end{abstract}
\end{titlepage}
\setcounter{page}{2}
Ultra-relativistic heavy-ion collisions create a system of deconfined
quarks and gluons, called the Quark--Gluon Plasma
(QGP)~\cite{Adams:2005dq,Gyulassy:2004zy,Andronic:2017pug} and provide the opportunity to study its properties.
In collisions with non-zero impact parameter, a large angular momentum of $O(10^{7})$ $\hslash$~\cite{Becattini:2007sr} and magnetic field of $O(10^{14})$ T \cite{Kharzeev:2007jp} are also expected. While the
magnetic field is short lived (a few fm/$c$), the
angular momentum is conserved and could affect the
system throughout its evolution. Experimental observables like correlations in azimuthal angle~\cite{Fries:2017ina,Voronyuk:2011jd} could be used to study the influence of these initial conditions on the properties and the dynamical evolution of the QGP and its subsequent hadronization.

Spin-orbit interactions have wide observable consequences in several branches of
physics~\cite{Jdjackson,landau,Mayer:1949pd}. In the presence of a large angular momentum, the
spin-orbit coupling of quantum chromodynamics (QCD) could 
lead to a polarization of quarks followed by a net-polarization of vector mesons (\kst~and
\ph)~\cite{Voloshin:2004ha,Liang:2004ph,Liang:2004xn,Liang:2007ma,Yang:2017sdk} along the direction of
the angular momentum.

The spin state of a vector meson is described by a 3 $\times$ 3
Hermitian spin-density matrix~\cite{Yang:2017sdk}. 
Its trace is 1 and $\mathrm{\rho_{11}}$ and $\mathrm{\rho_{-1-1}}$ cannot be measured separately in two-body decays to pseudoscalar mesons. Consequently, there is only one independent diagonal element, \rh. The elements of the spin-density matrix can be studied by measuring the angular distributions of the decay products of vector mesons with respect to a
quantization axis. Here two different quantization axes are used: i) a vector perpendicular to the production
plane (PP) of the vector meson and ii) the normal to the reaction plane (RP) of the
system. The PP is defined by the flight direction of the vector meson and the
beam direction.  

The spin-density matrix element \rh~is determined from the distribution of the angle $\theta^{*}$ between the kaon decay daughter and the quantization 
axis in the decay rest frame~\cite{Schilling:1969um},[~\ref{sup_mat}],
\begin{equation}
\frac{\mathrm{d}N}{\mathrm{d}\cos{\theta^{*}}} \propto [ 1 - \rho_{00} + \cos^{2}{\theta^{*}}(3\rho_{00} - 1 ) ].
\label{eqn1}
\end{equation}
\rh~is 1/3 in the absence of spin alignment and the angular distribution in Eq.~\ref{eqn1} is uniform. 
The experimental signature of spin alignment is a non-uniform angular distribution (\rh~$\neq$ 1/3).

The direction of the angular momentum in non-central heavy-ion collisions is perpendicular to the reaction plane (subtended by the
beam axis and impact parameter)~\cite{Liang:2004ph}. The spin-orbit interaction is expected to lead to spin alignment with respect to the reaction plane (RP). The reaction plane orientation cannot be measured directly, but is estimated from the final state distributions of particles. This experimentally measured plane is called the event plane (EP)~\cite{Poskanzer:1998yz}. 
The deviation of the EP with respect to the RP is corrected using the EP resolution ($R$) and observed $\rho^{obs}_{00}$~\cite{Tang:2018qtu},
\begin{equation}
\rho_{00} = \frac{1}{3} + \left (\rho^{\mathrm{obs}}_{00} - \frac{1}{3}\right)\frac{4}{1+3R}.
\label{eqn2}
\end{equation}

There are specific qualitative predictions for the spin alignment
effect~\cite{Liang:2004xn}: (a) \rh~$>$ 1/3 if the hadronization of a polarized parton
proceeds via fragmentation and less than 1/3 for hadronization via
recombination, 
(b) \rh~is expected to have a smaller deviation from 1/3 for both central (impact parameter $\lesssim$ 3 fm) and peripheral (impact parameter $\gtrsim$ 11 fm) heavy-ion collisions, and a maximum deviation for mid-central collisions, where the angular momentum is also maximal, 
(c) the \rh~value is expected to have maximum deviation from 1/3 at low \pt~and
reach the value of 1/3 at high \pt~in the recombination scenario, and (d) the effect is
expected to be larger for \kst~compared to \ph~due to their constituent quark composition. The initial large magnetic field might also affect the \rh~values~\cite{Yang:2017sdk}. This leads to \rh~$>$ 1/3 for neutral and \rh~$<$ 1/3 for charged vector mesons. 
Hence magnetic field and angular momentum could have opposite effects on electrically neutral $K^{*0}$, $\phi$.
All of these features are probed for \kst~and~\ph~mesons in Pb--Pb collisions presented in this letter.
As a cross check, a control measurement is carried out using $pp$ collisions, which do not possess large
initial angular momentum, and the same analysis is done in \pb~collisions for \Kzs~meson, which has zero spin. 
In addition, the measurements are carried out by randomizing the directions of the event (RndEP) and production planes (RndPP).

The analyses are carried out using 43 million minimum bias $pp$ collisions at $\sqrt{s}$ = 13 TeV, taken in 2015 and 14 million minimum bias \pb~collisions at \snn~= 2.76 TeV, collected in 2010.  The minimum bias event selection in Pb--Pb collisions require at least one hit in any of V0A, V0C, and Silicon Pixel Detectors while in $pp$ collisions at least one hit in both V0A and V0C is required.
The events are further required to have a primary vertex position within $\pm$~10 cm of the detector center along the beam axis. 
The events were classified by collision centrality classes based on the amplitude measured in the V0 counters~\cite{Abelev:2013qoq}. The measurements are performed at midrapidity ($|y| <$ 0.5) as a function of \pt~and are reported for $pp$ collisions as well as for different centrality classes in \pb~collisions. The \Kzs~analysis is performed only for \pb~collisions in the 20--40\% centrality class which corresponds to the top 20--40$\%$ of V0 amplitude distribution. The details of the ALICE detector, trigger conditions, centrality selection, and second order event plane estimation using the V0 detectors at forward rapidity, can be found in~\cite{Aamodt:2008zz,Aamodt:2010cz,Abelev:2013qoq,Abelev:2014pua}. The \kst~and \ph~candidates are reconstructed via their decays into charged $K$$\pi$ and $KK$ pairs, respectively, while the \Kzs~is reconstructed via its decay into two charged pions. The Time Projection Chamber (TPC)~\cite{Alme:2010ke} and Time-of-Flight (TOF) detector~\cite{Dellacasa:2000kh} are used to identify the decay products of these mesons via specific ionization energy loss and time-of-flight measurements, respectively. The \kst~and \ph~yields are determined via the invariant mass technique~\cite{Adam:2017zbf,Abelev:2014uua,Abelev:2013xaa}. The background coming from combinatorial pairs and misidentified particles is removed by constructing the invariant mass distribution from so-called mixed events for the \kst~and \ph~\cite{Adam:2017zbf,Abelev:2014uua}. The combinatorial background for the \Kzs~candidates is significantly reduced by selecting the distinctive V-shaped decay topology~\cite{Abelev:2013xaa}.

\begin{figure}
      \begin{center}
        \includegraphics[scale=0.42]{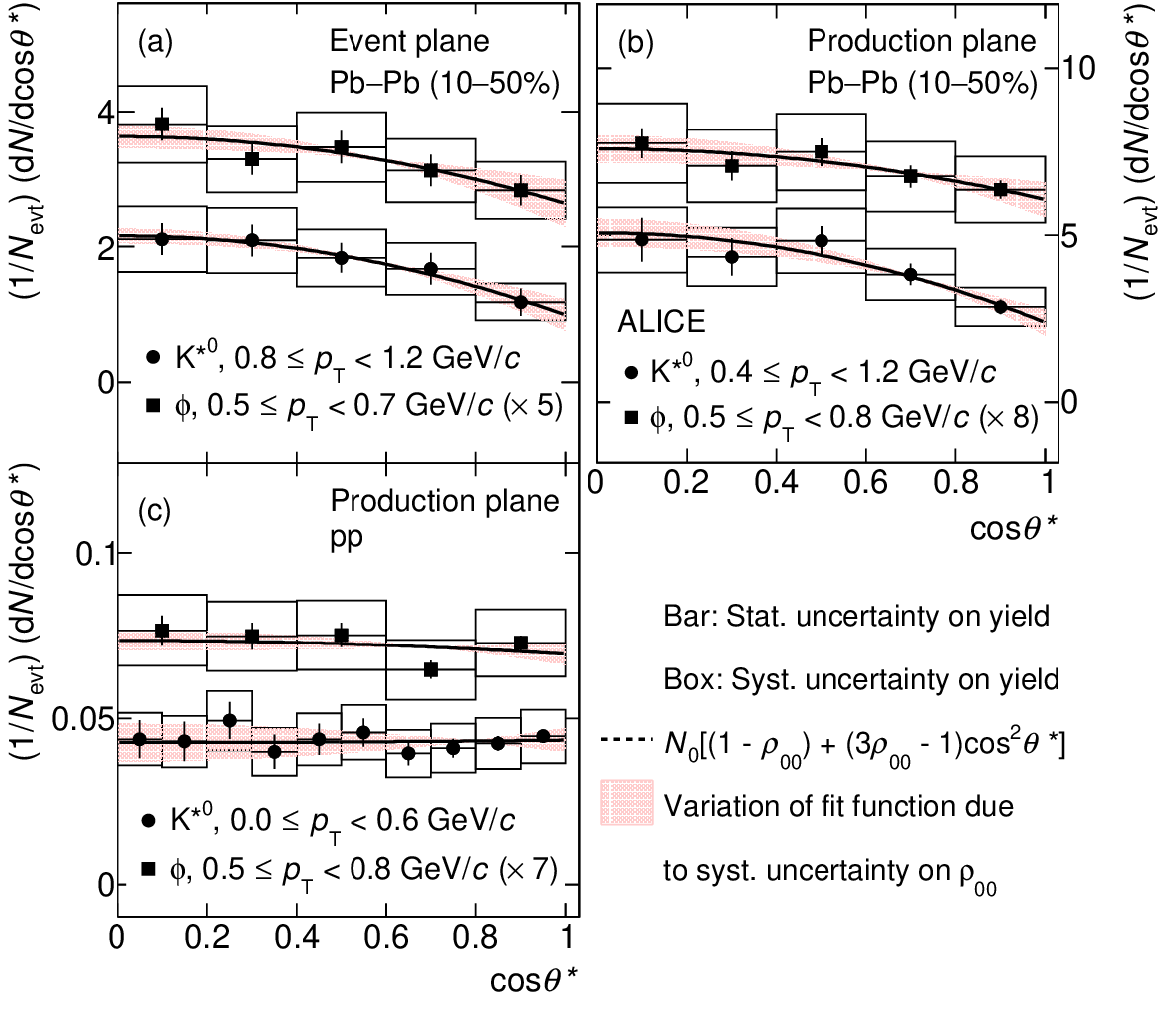}
       \caption{(Color online) Angular distribution of the decay
         daughter in the rest frame of the meson with respect to the
         quantization axis at $|y|~<$ 0.5 for $pp$ collisions at \s~= 13 TeV and \pb~collisions at \snn~= 2.76
         TeV. Panels (a) and (b) show results for \kst~and \ph~with respect to EP and PP. Panel (c) is the results for vector mesons in $pp$ collisions with respect to PP.
}
        \label{dist}
\vspace{-0.8cm}
      \end{center}
    \end{figure}

The invariant mass distributions are fitted with a Breit-Wigner and Voigtian (convolution of Breit-Wigner and Gaussian distributions) 
function for the \kst~and \ph~signals, respectively, along with a $2^{\mathrm{nd}}$ order polynomial that describes the residual background~\cite{Adam:2017zbf,Abelev:2014uua}.
Extracted yields are then corrected for the reconstruction efficiency and acceptance in each $\cos{\theta^{*}}$ and \pt~bin~\cite{Adam:2017zbf,Abelev:2014uua}. The reconstruction efficiency is determined from Monte Carlo simulations of the ALICE detector response based on GEANT3 simulation~\cite{Adam:2017zbf,Abelev:2014uua}. The signal extraction procedures for the vector mesons and \Kzs~are identical to those used in earlier publications reporting the \pt~distribution of the mesons~\cite{Adam:2017zbf,Abelev:2014uua,Abelev:2013xaa}. The mass peak positions and widths of the resonances across all the $\cos{\theta^{*}}$ bins for various \pt~intervals in $pp$ collisions and in different centrality classes of Pb--Pb collisions are consistent with those obtained from earlier analyses~\cite{Adam:2017zbf,Abelev:2014uua,Abelev:2013xaa} and no significant dependence on $\cos{\theta^{*}}$ is seen. 
The resulting efficiency and acceptance corrected
$\mathrm{d}N/\mathrm{d}\cos{\theta^{*}}$ distributions for selected
\pt~intervals in minimum bias $pp$ collisions and in 10--50\% central
\pb~collisions are shown in Fig.~\ref{dist}. These distributions are
fitted with the functional form given in Eq.~\ref{eqn1} to determine
$\rho_{00}$ for each \pt~bin in $pp$ and \pb~collisions. For the EP
results, the resolution values $R$ are 0.71, 0.53, 0.72, 0.66, and 0.40 for 10--50\%, 0--10\%, 10--30\%, 30--50\%, and 50--80\% collision centralities, respectively~\cite{alice-performance}. 

There are three main sources of systematic uncertainties in the measurements of the angular distribution of vector meson decays. (a) Meson yield extraction: this contribution is estimated by varying the fit ranges for the yield extraction, the normalization range for the signal+background and background invariant mass distributions, the procedure to integrate the signal function to get the yields, and by leaving the width of the resonance peak free or keeping it fixed to the PDG value as discussed in Ref.~\cite{Adam:2017zbf,Abelev:2014uua}.
The uncertainties for \rh~is at a level of 12(8)\% at the lowest \pt~and decrease with \pt~to 4(3)\% at the highest \pt~studied for the \kst(\ph). (b) Track selection: this contribution includes variations of the selection on the distance of closest approach to the collision vertex, the number of crossed pad rows in the TPC~\cite{Alme:2010ke}, the ratio of found clusters to the expected clusters, and the quality of the track fit. The systematic uncertainties for \rh~are 14(6)\% at the lowest \pt~and about 11(5)\% at the highest \pt~for \kst(\ph). (c) Particle identification: this is evaluated by varying the particle identification criteria related to the TPC and 
TOF detectors. The corresponding uncertainty is 5(3)\% at the lowest
\pt~and about 4(4.5)\% at the highest \pt~studied for \kst(\ph). Systematic uncertainties due to different variations are considered as uncorrelated and the total systematic uncertainty on \rh~is obtained by adding all the contributions in quadrature.
Several consistency checks are carried out and details can be found in the Supplemental Material~\ref{sup_mat}. The final measurement is reported for the average yield of particles (\kst) and anti-particles (\kstba) as results for \kst~and \kstba~were consistent.
    \begin{figure}
      \begin{center}
        \includegraphics[scale=0.45]{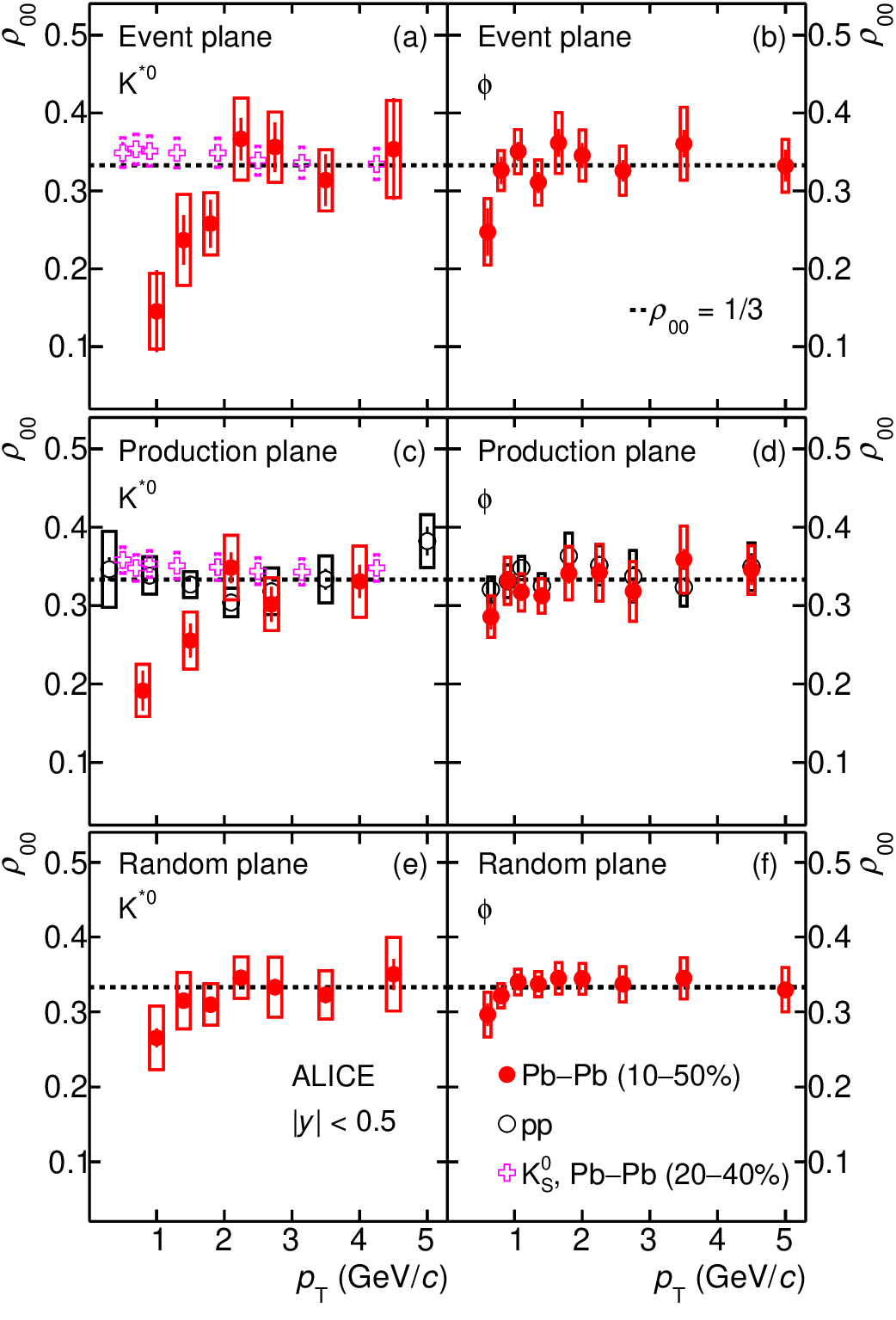}
       \caption{(Color online) 
       Transverse momentum dependence of \rh~for $K^{*0}$, \ph, and \Kzs~mesons at $|y| <$ 0.5 in Pb--Pb collisions at \snn~= 2.76 TeV and minimum bias $pp$ collisions at \s~= 13 TeV.
       Results are shown for spin alignment with respect to the event plane (panels a,b), production plane (c,d) and random event plane (e,f) for \kst~(left column) and \ph~(right column).
          The statistical and systematic uncertainties are 
          shown as bars and boxes, respectively.}
        \label{Fig:momentum}
        \vspace{-0.8cm}
      \end{center}
    \end{figure}

Figure~\ref{Fig:momentum} shows the measured \rh~as a function of \pt~for \kst~and \ph~mesons in $pp$ collisions and \pb~collisions, along with the measurements for \Kzs~in Pb--Pb collisions. 
In mid-central (10--50\%) \pb~collisions, \rh~is below 1/3 at the
lowest measured \pt~and increases to 1/3 within uncertainties
for \pt~$>$ 2 GeV/$c$. At low \pt, the central value of \rh~is smaller for \kst~than for \ph, although the results are compatible within uncertainties. In $pp$ collisions, \rh~is independent of \pt~and
equal to 1/3 within uncertainties. For the spin zero hadron
\Kzs, \rh~is consistent with 1/3 within uncertainties in \pb~collisions. 
The results with random event plane directions are also compatible with no spin alignment for the studied \pt~range, except for the smallest \pt~bin, where \rh~less than 1/3 but still larger than for EP and PP measurements. 
The results for the random production plane (the momentum vector direction
of each vector meson is randomized) are similar to RndEP measurements.
These results indicate that a spin alignment is present at lower \pt,
which is a qualitatively consistent with predictions~\cite{Liang:2004xn}.
    \begin{figure}
      \begin{center}
        \includegraphics[scale=0.42]{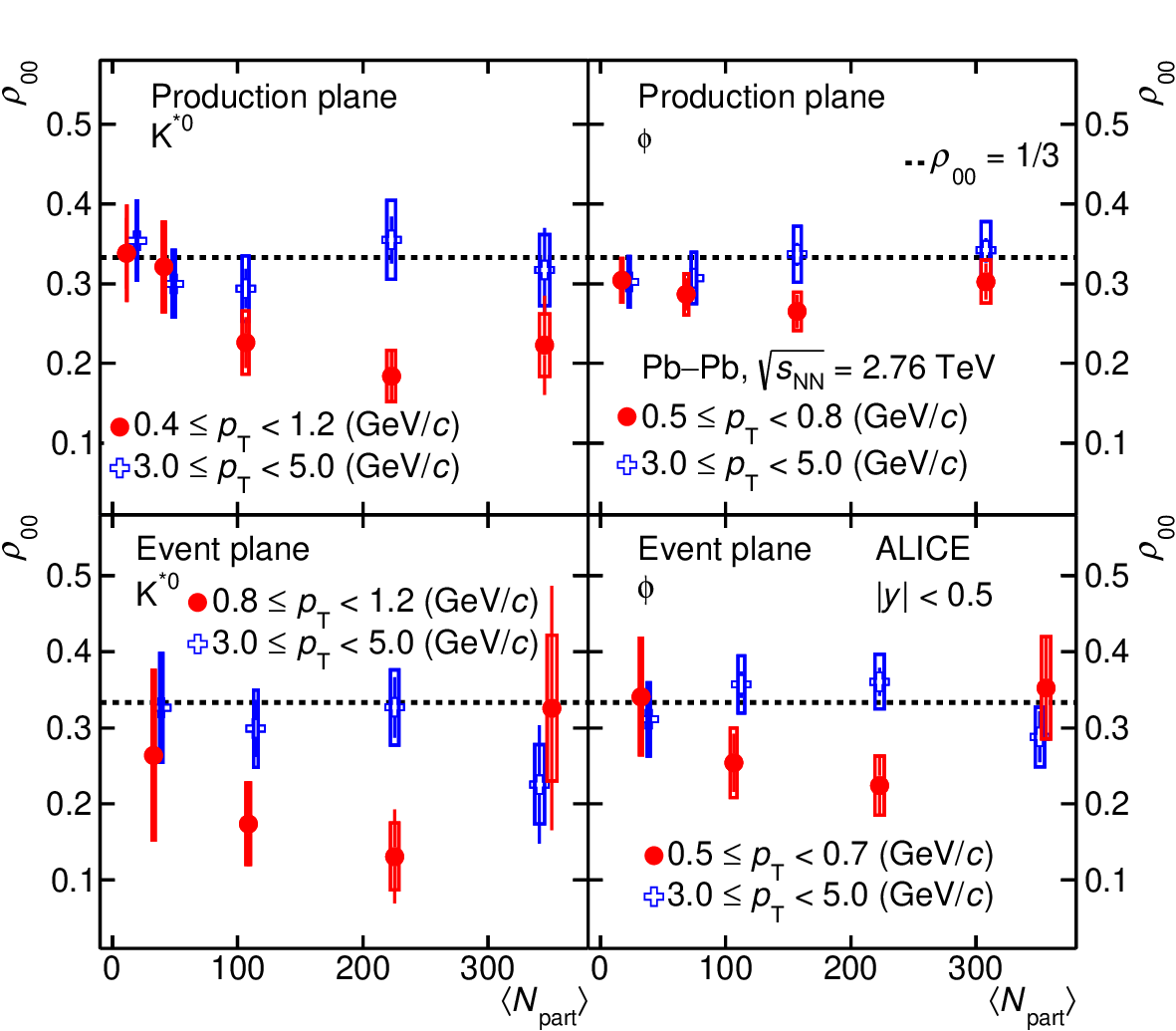}
       \caption{(Color online) Measurements of \rh~as a function of $\langle
         N_{\mathrm {part}} \rangle$ for \kst~and \ph~mesons at low
         and high \pt~in \pb~collisions. The statistical and systematic uncertainties are 
          shown as bars and boxes, respectively. Some data points are shifted horizontally for better visibility.}
        \label{Fig:Centrality}
        \vspace{-0.8cm}
      \end{center}
    \end{figure}

Figure~\ref{Fig:Centrality} shows \rh~for \kst~and \ph~mesons as a function of average number of participating
nucleons ($\langle N_{\mathrm {part}} \rangle$)~\cite{Abelev:2013qoq,Aamodt:2010cz}  for \pb~collisions at \snn~= 2.76~TeV. 
Large $\langle N_{\mathrm {part}} \rangle$ correspond to central collisions and small $\langle N_{\mathrm {part}} \rangle$ correspond to peripheral collisions (See Table~\ref{Tab:centrality_npart} of Supplemental Material~\ref{sup_mat}).
In the lowest \pt~range, \rh~shows maximum deviation from 1/3 for intermediate centrality and approaches 1/3 for both central and peripheral collisions. This centrality  dependence
is qualitatively consistent with the dependence of the initial angular momentum on
impact parameter in  
heavy-ion collisions ~\cite{Becattini:2007sr}.
At higher \pt, \rh~is consistent with 1/3 for all centrality classes. 
For the low-\pt~measurements in 10--30\% (20--40\% for \ph~meson w.r.t. PP) mid-central Pb--Pb collisions, the maximum deviations of \rh~from 1/3 with respect to the PP (EP) are 3.2 (2.6) $\sigma$ and 2.1 (1.9) $\sigma$ for \kst~and \ph~mesons, respectively.
The errors ($\sigma$) are calculated by adding statistical and systematic uncertainties in quadrature.

The relation between the $\mathrm{\rho_{00}}$ values with respect to different quantization axes can be expressed using Eq.~\ref{eqn2} and calculating the corresponding factor $R$. This gives $\Delta\mathrm{\rho_{00}(RndEP)}=\Delta\mathrm{\rho_{00}(EP)}\times\frac{1}{4}$ ($R$ = 0 for random plane) and $\Delta\mathrm{\rho_{00}(PP)}=\Delta\mathrm{\rho_{00}(EP)}\times\frac{1+3v_{2}}{4}$ ($R$ = $v_{2}$ for production plane, where $v_{2}$ is the second Fourier coefficient of the azimuthal distribution of produced particles relative to the event plane angle). Here $\Delta$$\mathrm{\rho_{00}}$ = $\mathrm{\rho_{00}}$ - 1/3. This is further confirmed using a toy model simulation with the PYTHIA 8.2 event generator~\cite{Sjostrand:2014zea} by incorporating $v_{2}$ and spin alignment (See Supplemental Material~\ref{sup_mat} for further details).

%
In the past, spin alignment measurements in $\mathrm{e}^{+}\mathrm{e}^{-}$~\cite{Ackerstaff:1997kj,Ackerstaff:1997kd,Abreu:1997wd}, hadron--proton~\cite{Barth:1982td} and nucleon--nucleus
collisions~\cite{Aleev:2000tb} were carried out to understand the role of spin in the dynamics of particle production, finding \rh~$>$ 1/3 and off-diagonal elements close to
zero with respect to the PP. For $pp$ collisions at $\sqrt{s}$ = 13 TeV, we find \rh~$\sim$ 1/3 within the studied \pt~range (see Fig.~\ref{Fig:momentum}). 
New preliminary results from RHIC have found deviations of \rh~from 1/3 indicating spin alignment for vector mesons at lower \snn~\cite{Zhou:2019lun, Abelev:2008ag}.
The \rh~for \ph~mesons in mid-central Pb--Pb collisions at \snn~= 2.76 TeV is  less than 1/3 while the preliminary finding for mid-central Au--Au collisions at \snn = 200 GeV is \rh~greater than 1/3.  The \rh~$>$ 1/3 for \ph~mesons has been interpreted as evidence for a coherent \ph~meson field~\cite{Sheng:2019kmk}. 
Similar conclusions cannot be easily applied to \kst~as it consists of valence quarks of unequal mass (s and $\bar{\rm{d}}$), which makes it impossible to separate the effects of vorticity and due to electromangetic and mesonic fields.
Significant polarization of $\Lambda$ baryons (spin = 1/2) was reported at
low RHIC energies. The polarization is found to decrease with increasing \snn~\cite{STAR:2017ckg,Adam:2018ivw}. At the LHC,
the global polarization for $\Lambda$ baryon is compatible with zero within uncertainties ($P_{\Lambda}$ (\%) = 0.01 $\pm$ 0.06 $\pm$ 0.03)~\cite{alice-gp}.
The spin alignment for vector mesons in heavy ion collisions could have contributions from angular momentum~\cite{Liang:2004ph,Liang:2004xn}, electromagnetic fields~\cite{Yang:2017sdk} and mesonic fields~\cite{Sheng:2019kmk}. While no quantitative theoretical calculation for vector meson polarization at LHC energies exists, the expected order of magnitude can be estimated and the measurements for vector mesons and hyperons can be related in a model dependent way. 
Considering only the angular momentum contribution and recombination as the process of hadronization~\cite{Liang:2004xn}, the \rh~of vector mesons are related to quark polarization as $\rh = \frac{1-P_{q}P_{\bar q}}{3+P_{q}P_{\bar q}}$ where $P_{q}$ and $P_{\bar q}$ are quark and anti-quark polarization, respectively. 
Assuming $P_{u}$ = $P_{\bar u}$ = $P_{d}$ = $P_{\bar d}$ and $P_{s}$ = $P_{\bar s}$, the measured \pt~integrated \rh~values for \kst~and \ph~mesons in 10--50$\%$ Pb--Pb collisions could translate to light quark polarization of $\sim$0.8 and strange quark polarization of $\sim$0.2.
Using a thermal and non-relativistic approach as discussed in~\cite{Becattini:2016gvu}, vorticity ($\omega$) and temperature ($T$) are related to \rh~and vector meson polarization ($P_{\rm{V}}$) as $\rh \simeq\frac{1}{3}\left(1-\frac{\left( \omega/T\right)^{2}}{3}\right)$ and $P_{\rm{V}}$ $\simeq$ (2$\omega$/3$T$), respectively. 
Also in this approach, the measured \rh~for \kst~would correspond to \kst~polarization of $\sim$0.6 and the \rh~for \ph~mesons would give \ph~meson polarization of $\sim$0.3. 

In the recombination model, $\Lambda$ polarization depends linearly on quark polarization whereas vector meson polarization depends quadratically on it. One would therefore expect the polarization for \kst~to be of the same order or smaller than the one measured for the $\Lambda$ at LHC~\cite{alice-gp}, i.e. vanishing small ($O$( 0.01\% )) rather than order 1. The large effect observed for the \rh~in mid-central Pb--Pb collisions at low \pt~is therefore puzzling.  This result should stimulate further theoretical work in order to study which effects could make such a huge difference between $\Lambda$ and \kst~polarization. Possible reasons may include the transfer of the quark polarization to the hadrons (baryon vs. meson), details of the hadronization mechanism (recombination vs. fragmentation), re-scattering, regeneration, and possibly the lifetime and mass of the relevant  hadron. Moreover, the vector mesons are predominantly directly produced whereas the hyperons have large contributions from resonance decays.
 
In conclusion, for the first time, evidence has been found for a significant spin alignment of vector mesons in heavy-ion collisions.
The effect is strongest at low \pt~with respect to a vector perpendicular to the reaction plane and for mid-central (10--50\%) collisions. These observations are qualitatively consistent with expectations from the effect of large initial angular momentum in non-central heavy-ion collisions, which leads to quark polarization via spin-orbit coupling, subsequently transferred to hadronic degrees of freedom by hadronization via recombination. However, the measured spin alignment is surprisingly large compared to the polarization measured for $\Lambda$ hyperons where, in addition, a strong decrease in polarization with \snn~is observed. 
In future measurements, the difference in the polarization of $K^{*\pm}$ and $K^{*o}$, due to their difference in magnetic moment, would be directly sensitive to the effect of the large initial magnetic field produced in heavy-ion collisions.

    \newenvironment{acknowledgement}{\relax}{\relax}
    \begin{acknowledgement}
      \section*{Acknowledgements}

The ALICE Collaboration would like to thank all its engineers and technicians for their invaluable contributions to the construction of the experiment and the CERN accelerator teams for the outstanding performance of the LHC complex.
The ALICE Collaboration gratefully acknowledges the resources and support provided by all Grid centres and the Worldwide LHC Computing Grid (WLCG) collaboration.
The ALICE Collaboration acknowledges the following funding agencies for their support in building and running the ALICE detector:
A. I. Alikhanyan National Science Laboratory (Yerevan Physics Institute) Foundation (ANSL), State Committee of Science and World Federation of Scientists (WFS), Armenia;
Austrian Academy of Sciences, Austrian Science Fund (FWF): [M 2467-N36] and Nationalstiftung f\"{u}r Forschung, Technologie und Entwicklung, Austria;
Ministry of Communications and High Technologies, National Nuclear Research Center, Azerbaijan;
Conselho Nacional de Desenvolvimento Cient\'{\i}fico e Tecnol\'{o}gico (CNPq), Financiadora de Estudos e Projetos (Finep), Funda\c{c}\~{a}o de Amparo \`{a} Pesquisa do Estado de S\~{a}o Paulo (FAPESP) and Universidade Federal do Rio Grande do Sul (UFRGS), Brazil;
Ministry of Education of China (MOEC) , Ministry of Science \& Technology of China (MSTC) and National Natural Science Foundation of China (NSFC), China;
Ministry of Science and Education and Croatian Science Foundation, Croatia;
Centro de Aplicaciones Tecnol\'{o}gicas y Desarrollo Nuclear (CEADEN), Cubaenerg\'{\i}a, Cuba;
Ministry of Education, Youth and Sports of the Czech Republic, Czech Republic;
The Danish Council for Independent Research | Natural Sciences, the VILLUM FONDEN and Danish National Research Foundation (DNRF), Denmark;
Helsinki Institute of Physics (HIP), Finland;
Commissariat \`{a} l'Energie Atomique (CEA), Institut National de Physique Nucl\'{e}aire et de Physique des Particules (IN2P3) and Centre National de la Recherche Scientifique (CNRS) and R\'{e}gion des  Pays de la Loire, France;
Bundesministerium f\"{u}r Bildung und Forschung (BMBF) and GSI Helmholtzzentrum f\"{u}r Schwerionenforschung GmbH, Germany;
General Secretariat for Research and Technology, Ministry of Education, Research and Religions, Greece;
National Research, Development and Innovation Office, Hungary;
Department of Atomic Energy Government of India (DAE), Department of Science and Technology, Government of India (DST), University Grants Commission, Government of India (UGC) and Council of Scientific and Industrial Research (CSIR), India;
Indonesian Institute of Science, Indonesia;
Centro Fermi - Museo Storico della Fisica e Centro Studi e Ricerche Enrico Fermi and Istituto Nazionale di Fisica Nucleare (INFN), Italy;
Institute for Innovative Science and Technology , Nagasaki Institute of Applied Science (IIST), Japanese Ministry of Education, Culture, Sports, Science and Technology (MEXT) and Japan Society for the Promotion of Science (JSPS) KAKENHI, Japan;
Consejo Nacional de Ciencia (CONACYT) y Tecnolog\'{i}a, through Fondo de Cooperaci\'{o}n Internacional en Ciencia y Tecnolog\'{i}a (FONCICYT) and Direcci\'{o}n General de Asuntos del Personal Academico (DGAPA), Mexico;
Nederlandse Organisatie voor Wetenschappelijk Onderzoek (NWO), Netherlands;
The Research Council of Norway, Norway;
Commission on Science and Technology for Sustainable Development in the South (COMSATS), Pakistan;
Pontificia Universidad Cat\'{o}lica del Per\'{u}, Peru;
Ministry of Science and Higher Education and National Science Centre, Poland;
Korea Institute of Science and Technology Information and National Research Foundation of Korea (NRF), Republic of Korea;
Ministry of Education and Scientific Research, Institute of Atomic Physics and Ministry of Research and Innovation and Institute of Atomic Physics, Romania;
Joint Institute for Nuclear Research (JINR), Ministry of Education and Science of the Russian Federation, National Research Centre Kurchatov Institute, Russian Science Foundation and Russian Foundation for Basic Research, Russia;
Ministry of Education, Science, Research and Sport of the Slovak Republic, Slovakia;
National Research Foundation of South Africa, South Africa;
Swedish Research Council (VR) and Knut \& Alice Wallenberg Foundation (KAW), Sweden;
European Organization for Nuclear Research, Switzerland;
Suranaree University of Technology (SUT), National Science and Technology Development Agency (NSDTA) and Office of the Higher Education Commission under NRU project of Thailand, Thailand;
Turkish Atomic Energy Agency (TAEK), Turkey;
National Academy of  Sciences of Ukraine, Ukraine;
Science and Technology Facilities Council (STFC), United Kingdom;
National Science Foundation of the United States of America (NSF) and United States Department of Energy, Office of Nuclear Physics (DOE NP), United States of America.   
      
    \end{acknowledgement}
    \bibliographystyle{utphys}
    \bibliography{Spin_alignment_prl}
    
    \newpage
    \appendix

    \section{The ALICE Collaboration}
    \label{app:collab}

\begingroup
\small
\begin{flushleft}
S.~Acharya\Irefn{org141}\And 
D.~Adamov\'{a}\Irefn{org94}\And 
A.~Adler\Irefn{org74}\And 
J.~Adolfsson\Irefn{org80}\And 
M.M.~Aggarwal\Irefn{org99}\And 
G.~Aglieri Rinella\Irefn{org33}\And 
M.~Agnello\Irefn{org30}\And 
N.~Agrawal\Irefn{org10}\textsuperscript{,}\Irefn{org53}\And 
Z.~Ahammed\Irefn{org141}\And 
S.~Ahmad\Irefn{org16}
% ...
\And
\renewcommand\labelenumi{\textsuperscript{\theenumi}~}

\section*{Affiliation notes}
\renewcommand\theenumi{\roman{enumi}}
\begin{Authlist}
\item \Adef{org*}Deceased
\item \Adef{orgI}Dipartimento DET del Politecnico di Torino, Turin, Italy
\item \Adef{orgII}M.V. Lomonosov Moscow State University, D.V. Skobeltsyn Institute of Nuclear, Physics, Moscow, Russia
\item \Adef{orgIII}Department of Applied Physics, Aligarh Muslim University, Aligarh, India
\item \Adef{orgIV}Institute of Theoretical Physics, University of Wroclaw, Poland
\end{Authlist}

\section*{Collaboration Institutes}
\renewcommand\theenumi{\arabic{enumi}~}
\begin{Authlist}
\item \Idef{org1}A.I. Alikhanyan National Science Laboratory (Yerevan Physics Institute) Foundation, Yerevan, Armenia
\item \Idef{org2}Bogolyubov Institute for Theoretical Physics, National Academy of Sciences of Ukraine, Kiev, Ukraine
\item \Idef{org3}Bose Institute, Department of Physics  and Centre for Astroparticle Physics and Space Science (CAPSS), Kolkata, India
\item \Idef{org4}Budker Institute for Nuclear Physics, Novosibirsk, Russia
\item \Idef{org5}California Polytechnic State University, San Luis Obispo, California, United States
\item \Idef{org6}Central China Normal University, Wuhan, China
\item \Idef{org7}Centre de Calcul de l'IN2P3, Villeurbanne, Lyon, France
\item \Idef{org8}Centro de Aplicaciones Tecnol\'{o}gicas y Desarrollo Nuclear (CEADEN), Havana, Cuba
\item \Idef{org9}Centro de Investigaci\'{o}n y de Estudios Avanzados (CINVESTAV), Mexico City and M\'{e}rida, Mexico
\item \Idef{org10}Centro Fermi - Museo Storico della Fisica e Centro Studi e Ricerche ``Enrico Fermi', Rome, Italy
\item \Idef{org11}Chicago State University, Chicago, Illinois, United States
\item \Idef{org12}China Institute of Atomic Energy, Beijing, China
\item \Idef{org13}Comenius University Bratislava, Faculty of Mathematics, Physics and Informatics, Bratislava, Slovakia
\item \Idef{org14}COMSATS University Islamabad, Islamabad, Pakistan
\item \Idef{org15}Creighton University, Omaha, Nebraska, United States
\item \Idef{org16}Department of Physics, Aligarh Muslim University, Aligarh, India
\item \Idef{org17}Department of Physics, Pusan National University, Pusan, Republic of Korea
\item \Idef{org18}Department of Physics, Sejong University, Seoul, Republic of Korea
\item \Idef{org19}Department of Physics, University of California, Berkeley, California, United States
\item \Idef{org20}Department of Physics, University of Oslo, Oslo, Norway
\item \Idef{org21}Department of Physics and Technology, University of Bergen, Bergen, Norway
\item \Idef{org22}Dipartimento di Fisica dell'Universit\`{a} 'La Sapienza' and Sezione INFN, Rome, Italy
\item \Idef{org23}Dipartimento di Fisica dell'Universit\`{a} and Sezione INFN, Cagliari, Italy
\item \Idef{org24}Dipartimento di Fisica dell'Universit\`{a} and Sezione INFN, Trieste, Italy
\item \Idef{org25}Dipartimento di Fisica dell'Universit\`{a} and Sezione INFN, Turin, Italy
\item \Idef{org26}Dipartimento di Fisica e Astronomia dell'Universit\`{a} and Sezione INFN, Bologna, Italy
\item \Idef{org27}Dipartimento di Fisica e Astronomia dell'Universit\`{a} and Sezione INFN, Catania, Italy
\item \Idef{org28}Dipartimento di Fisica e Astronomia dell'Universit\`{a} and Sezione INFN, Padova, Italy
\item \Idef{org29}Dipartimento di Fisica `E.R.~Caianiello' dell'Universit\`{a} and Gruppo Collegato INFN, Salerno, Italy
\item \Idef{org30}Dipartimento DISAT del Politecnico and Sezione INFN, Turin, Italy
\item \Idef{org31}Dipartimento di Scienze e Innovazione Tecnologica dell'Universit\`{a} del Piemonte Orientale and INFN Sezione di Torino, Alessandria, Italy
\item \Idef{org32}Dipartimento Interateneo di Fisica `M.~Merlin' and Sezione INFN, Bari, Italy
\item \Idef{org33}European Organization for Nuclear Research (CERN), Geneva, Switzerland
\item \Idef{org34}Faculty of Electrical Engineering, Mechanical Engineering and Naval Architecture, University of Split, Split, Croatia
\item \Idef{org35}Faculty of Engineering and Science, Western Norway University of Applied Sciences, Bergen, Norway
\item \Idef{org36}Faculty of Nuclear Sciences and Physical Engineering, Czech Technical University in Prague, Prague, Czech Republic
\item \Idef{org37}Faculty of Science, P.J.~\v{S}af\'{a}rik University, Ko\v{s}ice, Slovakia
\item \Idef{org38}Frankfurt Institute for Advanced Studies, Johann Wolfgang Goethe-Universit\"{a}t Frankfurt, Frankfurt, Germany
\item \Idef{org39}Fudan University, Shanghai, China
\item \Idef{org40}Gangneung-Wonju National University, Gangneung, Republic of Korea
\item \Idef{org41}Gauhati University, Department of Physics, Guwahati, India
\item \Idef{org42}Helmholtz-Institut f\"{u}r Strahlen- und Kernphysik, Rheinische Friedrich-Wilhelms-Universit\"{a}t Bonn, Bonn, Germany
\item \Idef{org43}Helsinki Institute of Physics (HIP), Helsinki, Finland
\item \Idef{org44}High Energy Physics Group,  Universidad Aut\'{o}noma de Puebla, Puebla, Mexico
\item \Idef{org45}Hiroshima University, Hiroshima, Japan
\item \Idef{org46}Hochschule Worms, Zentrum  f\"{u}r Technologietransfer und Telekommunikation (ZTT), Worms, Germany
\item \Idef{org47}Horia Hulubei National Institute of Physics and Nuclear Engineering, Bucharest, Romania
\item \Idef{org48}Indian Institute of Technology Bombay (IIT), Mumbai, India
\item \Idef{org49}Indian Institute of Technology Indore, Indore, India
\item \Idef{org50}Indonesian Institute of Sciences, Jakarta, Indonesia
\item \Idef{org51}INFN, Laboratori Nazionali di Frascati, Frascati, Italy
\item \Idef{org52}INFN, Sezione di Bari, Bari, Italy
\item \Idef{org53}INFN, Sezione di Bologna, Bologna, Italy
\item \Idef{org54}INFN, Sezione di Cagliari, Cagliari, Italy
\item \Idef{org55}INFN, Sezione di Catania, Catania, Italy
\item \Idef{org56}INFN, Sezione di Padova, Padova, Italy
\item \Idef{org57}INFN, Sezione di Roma, Rome, Italy
\item \Idef{org58}INFN, Sezione di Torino, Turin, Italy
\item \Idef{org59}INFN, Sezione di Trieste, Trieste, Italy
\item \Idef{org60}Inha University, Incheon, Republic of Korea
\item \Idef{org61}Institut de Physique Nucl\'{e}aire d'Orsay (IPNO), Institut National de Physique Nucl\'{e}aire et de Physique des Particules (IN2P3/CNRS), Universit\'{e} de Paris-Sud, Universit\'{e} Paris-Saclay, Orsay, France
\item \Idef{org62}Institute for Nuclear Research, Academy of Sciences, Moscow, Russia
\item \Idef{org63}Institute for Subatomic Physics, Utrecht University/Nikhef, Utrecht, Netherlands
\item \Idef{org64}Institute of Experimental Physics, Slovak Academy of Sciences, Ko\v{s}ice, Slovakia
\item \Idef{org65}Institute of Physics, Homi Bhabha National Institute, Bhubaneswar, India
\item \Idef{org66}Institute of Physics of the Czech Academy of Sciences, Prague, Czech Republic
\item \Idef{org67}Institute of Space Science (ISS), Bucharest, Romania
\item \Idef{org68}Institut f\"{u}r Kernphysik, Johann Wolfgang Goethe-Universit\"{a}t Frankfurt, Frankfurt, Germany
\item \Idef{org69}Instituto de Ciencias Nucleares, Universidad Nacional Aut\'{o}noma de M\'{e}xico, Mexico City, Mexico
\item \Idef{org70}Instituto de F\'{i}sica, Universidade Federal do Rio Grande do Sul (UFRGS), Porto Alegre, Brazil
\item \Idef{org71}Instituto de F\'{\i}sica, Universidad Nacional Aut\'{o}noma de M\'{e}xico, Mexico City, Mexico
\item \Idef{org72}iThemba LABS, National Research Foundation, Somerset West, South Africa
\item \Idef{org73}Jeonbuk National University, Jeonju, Republic of Korea
\item \Idef{org74}Johann-Wolfgang-Goethe Universit\"{a}t Frankfurt Institut f\"{u}r Informatik, Fachbereich Informatik und Mathematik, Frankfurt, Germany
\item \Idef{org75}Joint Institute for Nuclear Research (JINR), Dubna, Russia
\item \Idef{org76}Korea Institute of Science and Technology Information, Daejeon, Republic of Korea
\item \Idef{org77}KTO Karatay University, Konya, Turkey
\item \Idef{org78}Laboratoire de Physique Subatomique et de Cosmologie, Universit\'{e} Grenoble-Alpes, CNRS-IN2P3, Grenoble, France
\item \Idef{org79}Lawrence Berkeley National Laboratory, Berkeley, California, United States
\item \Idef{org80}Lund University Department of Physics, Division of Particle Physics, Lund, Sweden
\item \Idef{org81}Nagasaki Institute of Applied Science, Nagasaki, Japan
\item \Idef{org82}Nara Women{'}s University (NWU), Nara, Japan
\item \Idef{org83}National and Kapodistrian University of Athens, School of Science, Department of Physics , Athens, Greece
\item \Idef{org84}National Centre for Nuclear Research, Warsaw, Poland
\item \Idef{org85}National Institute of Science Education and Research, Homi Bhabha National Institute, Jatni, India
\item \Idef{org86}National Nuclear Research Center, Baku, Azerbaijan
\item \Idef{org87}National Research Centre Kurchatov Institute, Moscow, Russia
\item \Idef{org88}Niels Bohr Institute, University of Copenhagen, Copenhagen, Denmark
\item \Idef{org89}Nikhef, National institute for subatomic physics, Amsterdam, Netherlands
\item \Idef{org90}NRC Kurchatov Institute IHEP, Protvino, Russia
\item \Idef{org91}NRC «Kurchatov Institute»  - ITEP, Moscow, Russia
\item \Idef{org92}NRNU Moscow Engineering Physics Institute, Moscow, Russia
\item \Idef{org93}Nuclear Physics Group, STFC Daresbury Laboratory, Daresbury, United Kingdom
\item \Idef{org94}Nuclear Physics Institute of the Czech Academy of Sciences, \v{R}e\v{z} u Prahy, Czech Republic
\item \Idef{org95}Oak Ridge National Laboratory, Oak Ridge, Tennessee, United States
\item \Idef{org96}Ohio State University, Columbus, Ohio, United States
\item \Idef{org97}Petersburg Nuclear Physics Institute, Gatchina, Russia
\item \Idef{org98}Physics department, Faculty of science, University of Zagreb, Zagreb, Croatia
\item \Idef{org99}Physics Department, Panjab University, Chandigarh, India
\item \Idef{org100}Physics Department, University of Jammu, Jammu, India
\item \Idef{org101}Physics Department, University of Rajasthan, Jaipur, India
\item \Idef{org102}Physikalisches Institut, Eberhard-Karls-Universit\"{a}t T\"{u}bingen, T\"{u}bingen, Germany
\item \Idef{org103}Physikalisches Institut, Ruprecht-Karls-Universit\"{a}t Heidelberg, Heidelberg, Germany
\item \Idef{org104}Physik Department, Technische Universit\"{a}t M\"{u}nchen, Munich, Germany
\item \Idef{org105}Politecnico di Bari, Bari, Italy
\item \Idef{org106}Research Division and ExtreMe Matter Institute EMMI, GSI Helmholtzzentrum f\"ur Schwerionenforschung GmbH, Darmstadt, Germany
\item \Idef{org107}Rudjer Bo\v{s}kovi\'{c} Institute, Zagreb, Croatia
\item \Idef{org108}Russian Federal Nuclear Center (VNIIEF), Sarov, Russia
\item \Idef{org109}Saha Institute of Nuclear Physics, Homi Bhabha National Institute, Kolkata, India
\item \Idef{org110}School of Physics and Astronomy, University of Birmingham, Birmingham, United Kingdom
\item \Idef{org111}Secci\'{o}n F\'{\i}sica, Departamento de Ciencias, Pontificia Universidad Cat\'{o}lica del Per\'{u}, Lima, Peru
\item \Idef{org112}St. Petersburg State University, St. Petersburg, Russia
\item \Idef{org113}Stefan Meyer Institut f\"{u}r Subatomare Physik (SMI), Vienna, Austria
\item \Idef{org114}SUBATECH, IMT Atlantique, Universit\'{e} de Nantes, CNRS-IN2P3, Nantes, France
\item \Idef{org115}Suranaree University of Technology, Nakhon Ratchasima, Thailand
\item \Idef{org116}Technical University of Ko\v{s}ice, Ko\v{s}ice, Slovakia
\item \Idef{org117}Technische Universit\"{a}t M\"{u}nchen, Excellence Cluster 'Universe', Munich, Germany
\item \Idef{org118}The Henryk Niewodniczanski Institute of Nuclear Physics, Polish Academy of Sciences, Cracow, Poland
\item \Idef{org119}The University of Texas at Austin, Austin, Texas, United States
\item \Idef{org120}Universidad Aut\'{o}noma de Sinaloa, Culiac\'{a}n, Mexico
\item \Idef{org121}Universidade de S\~{a}o Paulo (USP), S\~{a}o Paulo, Brazil
\item \Idef{org122}Universidade Estadual de Campinas (UNICAMP), Campinas, Brazil
\item \Idef{org123}Universidade Federal do ABC, Santo Andre, Brazil
\item \Idef{org124}University of Cape Town, Cape Town, South Africa
\item \Idef{org125}University of Houston, Houston, Texas, United States
\item \Idef{org126}University of Jyv\"{a}skyl\"{a}, Jyv\"{a}skyl\"{a}, Finland
\item \Idef{org127}University of Liverpool, Liverpool, United Kingdom
\item \Idef{org128}University of Science and Technology of China, Hefei, China
\item \Idef{org129}University of South-Eastern Norway, Tonsberg, Norway
\item \Idef{org130}University of Tennessee, Knoxville, Tennessee, United States
\item \Idef{org131}University of the Witwatersrand, Johannesburg, South Africa
\item \Idef{org132}University of Tokyo, Tokyo, Japan
\item \Idef{org133}University of Tsukuba, Tsukuba, Japan
\item \Idef{org134}Universit\'{e} Clermont Auvergne, CNRS/IN2P3, LPC, Clermont-Ferrand, France
\item \Idef{org135}Universit\'{e} de Lyon, Universit\'{e} Lyon 1, CNRS/IN2P3, IPN-Lyon, Villeurbanne, Lyon, France
\item \Idef{org136}Universit\'{e} de Strasbourg, CNRS, IPHC UMR 7178, F-67000 Strasbourg, France, Strasbourg, France
\item \Idef{org137}Universit\'{e} Paris-Saclay Centre d'Etudes de Saclay (CEA), IRFU, D\'{e}partment de Physique Nucl\'{e}aire (DPhN), Saclay, France
\item \Idef{org138}Universit\`{a} degli Studi di Foggia, Foggia, Italy
\item \Idef{org139}Universit\`{a} degli Studi di Pavia, Pavia, Italy
\item \Idef{org140}Universit\`{a} di Brescia, Brescia, Italy
\item \Idef{org141}Variable Energy Cyclotron Centre, Homi Bhabha National Institute, Kolkata, India
\item \Idef{org142}Warsaw University of Technology, Warsaw, Poland
\item \Idef{org143}Wayne State University, Detroit, Michigan, United States
\item \Idef{org144}Westf\"{a}lische Wilhelms-Universit\"{a}t M\"{u}nster, Institut f\"{u}r Kernphysik, M\"{u}nster, Germany
\item \Idef{org145}Wigner Research Centre for Physics, Budapest, Hungary
\item \Idef{org146}Yale University, New Haven, Connecticut, United States
\item \Idef{org147}Yonsei University, Seoul, Republic of Korea
\end{Authlist}
\endgroup
            
    \newpage
  \section{Supplemental material}
\label{sup_mat}
\subsection{Angular distribution of the decay products of the vector meson}
The complete angular distribution of the decay products of the vector meson is expressed as,
\begin{eqnarray}
\frac{\mathrm{d}N}{\mathrm{d}\cos{\theta^{*}}\mathrm{d}\varphi^{*}}  &\propto& [\cos^{2}{\theta^{*}}\rho_{00} + \sin^{2}{\theta}(\rho_{11}+\rho_{-1-1})/2 \nonumber \\
&&~-~\sin{2\theta}(\cos{\phi^{*}}\mathrm{Re}\rho_{10} - \sin{\phi^{*}}\mathrm{Im}\rho_{10})/\sqrt{2} \nonumber \\
&&~+~\sin{2\theta}(\cos{\phi^{*}}\mathrm{Re}\rho_{-10} + \sin{\phi^{*}}\mathrm{Im}\rho_{-10})/\sqrt{2} \nonumber \\
&&~-~\sin^{2}{\theta}(\cos{2\phi^{*}}\mathrm{Re}\rho_{1-1} - \sin{2\phi^{*}}\mathrm{Im}\rho_{1-1})]
\label{eqn0}
\end{eqnarray}
The angle denoted here as $\theta^{*}$ is that made by one of the decay daughters in the rest frame of the vector meson with respect to the quantization  axis and $\varphi^{*}$ is the corresponding azimuthal angle. This expression is obtained by applying parity symmetry of QCD, the unit trace condition, and integrating over the azimuthal angle. 

 \subsection{Consistency checks}
In order to verify the measurements, several consistency checks are carried out. Specifically, the yields of vector mesons are summed over $\cos{\theta^{*}}$ bins for each \pt~interval to obtain the \pt~distributions; these are found to be consistent within the statistical uncertainties with the published \pt~distributions in \pb~collisions ~\cite{Adam:2017zbf,Abelev:2014uua}. Similarly a closure test (comparison between generated and reconstructed angular distribution) is carried out for the Monte Carlo (MC) data which is used to obtain the reconstruction efficiencies for the mesons. Two different event generators are used to determine the reconstruction efficiency and the results are consistent. The effect of the shape of the \pt~distributions in the MC simulations is studied in detail and the impact on the \rh~measurement is found to be small. The dependence of the reconstruction efficiency for a $\cos{\theta^{*}}$ range on the azimuthal angle of vector meson ($\phi_{V}$) relative to the event plane angle ($\Psi$) is also studied. The reconstruction efficiencies obtained in a $\cos{\theta^{*}}$ range by integrating over $\phi_{V} - \Psi$ are similar to the efficiency obtained by averaging over the $\phi_{V} - \Psi$ bins. Data samples with two different magnetic field polarities in the experiment are separately analyzed and the $\cos{\theta^{*}}$ distributions are found to be consistent. In addition, the analysis is performed separately for positive (0 $<$ $y$ $<$ 0.5) and negative (--0.5~$<$~$y$~$<$ 0) rapidity and also for \kst~versus \kstba; the different samples are also consistent.

\subsection{Analytical relation between EP and PP}
The relation between measured \rh~with respect to two different frames of references is
 \begin{equation}
\rh(\mathrm{A})-\frac{1}{3}=\left(\rh(\mathrm{B})-\frac{1}{3}\right)\left(\frac{1}{4}+\frac{3}{4}\cos{2\psi}\right),
\label{eq5} 
\end{equation}
where frame A is obtained by rotating frame B by angle $\psi$. Averaging over angle $\psi$ gives,
  \begin{equation}
\rh(\mathrm{A})-\frac{1}{3}=\left(\rh(\mathrm{B})-\frac{1}{3}\right)\left(\frac{1}{4}+\frac{3}{4}\langle\cos{2\psi}\rangle\right).
\label{eq6} 
\end{equation}
The transition from the EP to PP is obtained by taking into account the elliptic flow of the vector meson which leads to
 \begin{equation}
\langle\cos{2\psi}\rangle=\frac{1}{2\pi}\int_{-\pi}^{\pi} \cos(2\psi)[1+2v_{2}\cos(2\psi)]d\psi=v_{2}.
\label{eq7} 
\end{equation}
Using Eq.~\ref{eq6} and Eq.~\ref{eq7}, analytical relation between EP and PP can be expressed as,
 \begin{equation}
\rh(\mathrm{PP})-\frac{1}{3}=\left(\rh(\mathrm{EP})-\frac{1}{3}\right)\left(\frac{1+3v_{2}}{4}\right).
\label{eq8} 
\end{equation}

\subsection{Toy model simulation to understand the relation between EP, PP and RndEP} 
The spin density matrix element \rh~for vector mesons is measured with respect to EP, PP and RndEP. The measured \rh~values for \kst~at  0.8 $<$ \pt~$<$ 1.2 GeV/$c$ with respect to different planes in heavy-ion collisions has the following ordering \rh(EP) $<$ \rh(PP) $<$ \rh(RndEP). To understand this ordering a toy model simulation is carried out by using PYTHIA (version 8.2) event generator~\cite{Sjostrand:2014zea} which has no azimuthal anisotropy and spin alignment. For this study we have taken event plane angle as zero, which corresponds impact parameter along x-axis. In order to find the relations between different frames, $v_{2}$ (0.15 $\pm$ 0.06, value expected for hadrons with mass similar to \kst~in \pb~collisions at \snn~=~2.76 TeV~\cite{Abelev:2014pua}) is introduced to \kst~by appropriate rotation of its momentum in azimuthal plane. The modified angle in azimuthal plane is calculated by solving the following equation
\begin{equation}
\phi_{0}=\phi+v_{2}\sin2\phi,
\label{eq3}
\end{equation}
where $\phi_{0}$ is azimuthal angle of a \kst~in absence of $v_{2}$ and for a given value of $v_{2}$, $\phi_{0}$ transforms to $\phi$.  
Then spin alignment (\rh~= 0.125, same as measured in data) is introduced with respect to event plane by rotating the momentum of decay daughters in \kst~rest frame by solving
\begin{equation}
\cos{\theta^{*}_{0}} = 3/2 \times [(1-\rho_{00}) \cos{\theta^{*}} + 1/3(3\rho_{00}-1)\cos^{3}\theta^{*}].
\label{eq4} 
\end{equation}
Here $\theta^{*}_{0}$ is the angle made by the decay daughter of \kst~with the quantization axis in absence of spin alignment. $\theta^{*}_{0}$ transforms to $\theta^{*}$ to introduce a given input value of \rh. In this study we assume that the $\phi^{*}$ is remain fixed during the rotation. 
With these modifications, calculations as in the experimental data is carried out. The results are shown in Fig.~\ref{toymodel} for two cases, with and without $v_{2}$. The result corresponding to event plane is correctly retrieved in the model. The model results for $v_{2}$ = 0, are same for production plane and random event plane. However with $v_{2}$ = 0.15, the \rh(PP)~value is lower and closer to data for PP. The toy model reproduces the hierarchy observed in the \rh~values for various planes as observed in data. 
The physical picture is that spin alignment with respect to the event plane is coupled to that in the production plane through the elliptic flow of the system. The \rh(RndEP) is lower than 1/3 as the quantization axis is always perpendicular to the beam axis, resulting in a residual effect. If the quantization axis is random in 3 dimension, then the residual effect is not present and the \rh~value is consistent with 1/3. 
   \begin{figure}[hbtp]
      \begin{center}
        \includegraphics[scale=0.4]{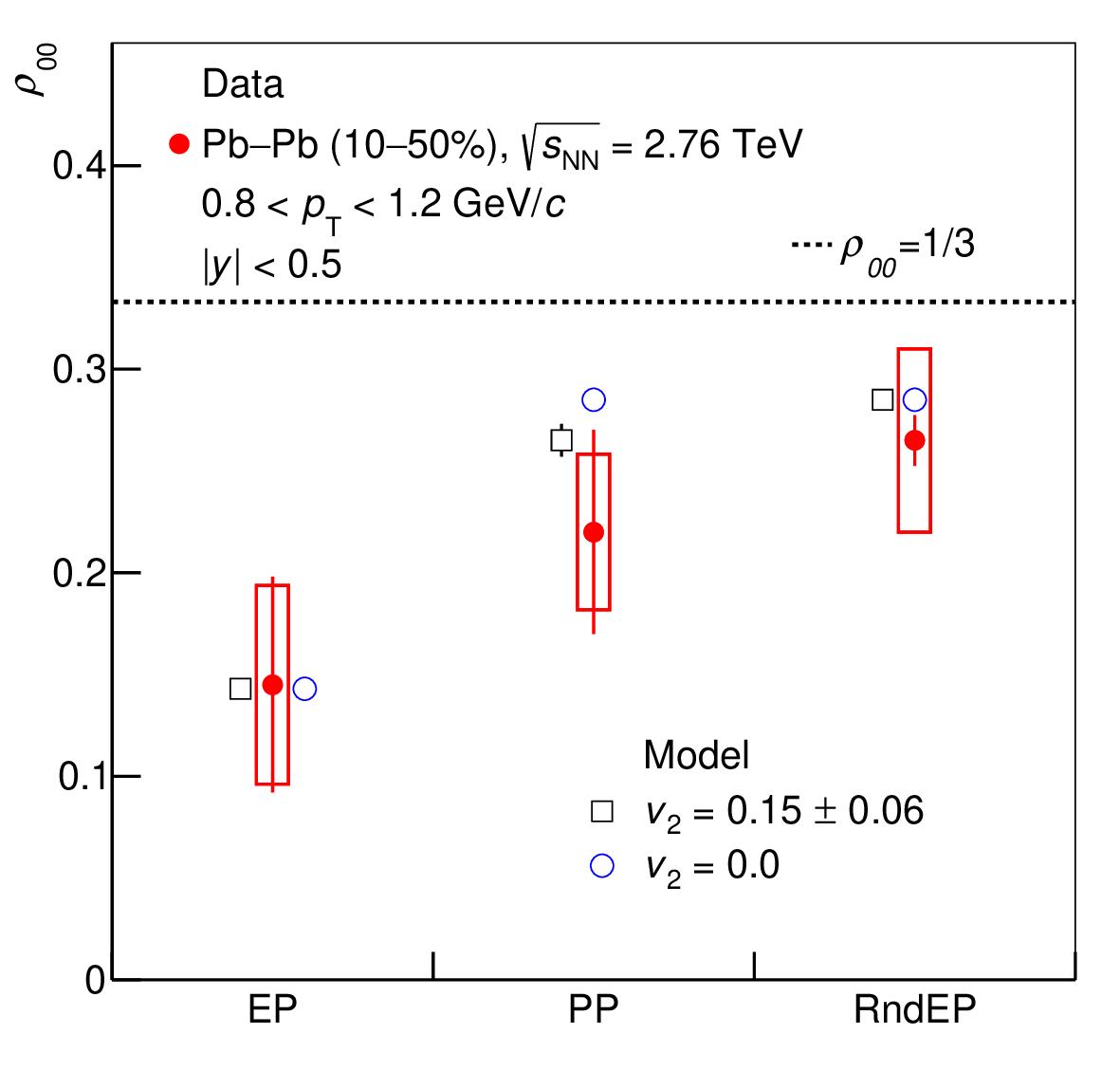}
       \caption{(Color online) \rh~values from data in 10--50\% Pb--Pb collisions at 0.8 $<$ \pt~$<$ 1.2 GeV/$c$ with respect to various planes compared with expectations from model simulations with and without added elliptic flow ($v_{2}$). The statistical and systematic uncertainties are shown as bars and boxes, respectively.} 
        \label{toymodel}
      \end{center}
    \end{figure}
    
    \subsection{$\langle N_{\mathrm {part}} \rangle$ values and centrality classes}
    Table~\ref{Tab:centrality_npart} shows collision centrality classes and their corresponding $\langle N_{\mathrm {part}} \rangle$ values, used in Fig.~\ref{Fig:Centrality}
    
    \begin{table}[hbtp]
     \caption{$\langle N_{\mathrm {part}} \rangle$ values for various centrality classes in Pb--Pb collisions at $\sqrt{s_{\mathrm{NN}}}$ = 2.76 TeV.}
     \label{Tab:centrality_npart}
      \begin{center}
       \scalebox{.8}{
      \begin{tabular}{|c|c|c|}
	\hline
	Particle & Centrality (\%) & $\langle N_{\mathrm {part}} \rangle$\\ 
	\hline
	\kst~(EP) and \ph~(EP)& 0--10\%&356.2 $\pm$ 3.8\\
	
	 &10--30\% &225.5 $\pm$ 3.5\\
	
	 &30--50\% &108.6 $\pm$ 1.8\\

	 &50--80\% & 32.9 $\pm$ 1.3\\
	\hline
	\kst~(PP) & 2--10\% & 345.3 $\pm$ 4.0\\
	
	 &10--30\% & 225.5 $\pm$ 3.5\\
	
	 &30--50\% & 108.6 $\pm$ 1.8\\

	&50--80\% & 32.9 $\pm$ 1.3 \\
	
	&50--70\% & 41.0 $\pm$ 1.3 \\
	
	&70-90\% & 11.4 $\pm$ 0.3 \\
	\hline
	\ph~(PP) & 0--20\% & 308.1 $\pm$ 3.7\\
	
	&20--40\% & 157.2 $\pm$ 3.1 \\

	&40--60\% & 68.6 $\pm$ 2.0 \\

	&60--80\% & 16.9 $\pm$ 0.5 \\
	\hline
	\end{tabular}}
	\end{center}
	\end{table}

\end{document}